# Do interests affect grant application success?
# The role of organizational proximity[1]

Charlie S Mom*, Peter van den Besselaar**,[2]

\* TMC, Amsterdam
charlie@teresamom.com

\*\* Vrije Universiteit Amsterdam, the Netherlands
p.a.a.vanden.besselaar@vu.nl

**Abstract:** Bias in grant allocation is a critical issue, as the expectation is that grants are given to the best researchers, and not to applicants that are socially, organizationally, or topic-wise near-by the decision-makers. In this paper, we investigate the effect of organizational proximity, defined as an applicant with the same affiliation as one of the panel members (a near-by panelist), on the probability of getting a grant. This study is based on one of the most prominent grant schemes in Europe, with overall excellent scientists as panel members. Various aspects of this organizational proximity are analyzed: Who gains from it? Does it have a gender dimension? Is it bias, or can it be explained by performance differences?

We do find that the probability to get funded increases significantly for those that apply in a panel where there is a panelist from the institution where the applicant has agreed to use the grant. At the same time, the effect differs between disciplines and countries, and men profit more of it than women do. Finally, depending on how one defines what counts as the best researchers, the near-by panelist effect can be interpreted as preferential attachment (quality links to quality) or as bias and particularism.

**Keywords:** Interest representation; conflict of interest; cronyism; nepotism; particularism; favoritism; proximity; grant selection; peer review; research funding; European Research Council; ERC.

---

[1] This paper is a result of the GendERC project, which was supported by the ERC (grant 610706). The authors thank Ulf Sandstrom (FPS, Stockholm) for comments a previous version, and Lucia Polo Alvarez (Tecnalia, Bilbao, Spain) for collecting data about the institutional affiliations of the panel members. Reviewers and participants of the 1st PEERE conference (Rome March 2018) provided useful comments that helped to improve the paper, as did the reviewers for the STI 2018 conference in Leiden, The Netherlands. Frédérique Sachwald (OST, HCERES, Paris) was helpful in explaining the complex French situation. In version V2 the theoretical section has been rewritten. Version V3 contains some minor editorial changes.

[2] Corresponding author

**Introduction**

From a Mertonian perspective, grant and career decisions in academia should be based on merit (Merton 1942). Deviation from this principle results in *particularism* or *favoritism*, which means that other than scholarly qualifications and performance play a role in the decision, such as gender, age, nationality, or characteristics like social, professional, or disciplinary network relations. In those cases, not the best and the most qualified get the job or the grant, but someone who is in some dimension close to the (preferences of) decision-makers.

Many studies have been done to answer the question whether grants are given to the best applicants, showing that the granted applicants on average have higher past performance than non-granted (Van Leeuwen & Moed 2012). But when the granted applicants were compared with a similarly large group of best-performing rejected applicants, the picture changed: the non-granted were on average at least equally good as the granted, resulting in high numbers of false positives and false negatives (Van den Besselaar & Leydesdorff 2009; Bornmann et al. 2010; Hornbostel et al. 2009). Furthermore, studies on the predictive validity of grant decisions also suggest that merit is not the main (or only) criterion for awarding grants (Van den Besselaar & Sandström 2015; Gallo et al. 2016; Wang et al 2019). If rewarding merit is not the modus operandi of panel review, then what are the mechanisms?

*Particularism in grant allocation*

Cole (1992) proposed a useful differentiation of the concept of particularism. At the one hand he distinguishes what could be called *social* particularism: selection processes based on social relations such as friendship, membership of political parties, being colleagues (cronyism), of on family ties (nepotism). This type of particularism is clearly against the Mertonian principle of universalism, although even here some reservation is needed. For example, if the science system is highly stratified, and top performers are concentrated in a small number of universities, attributed merit and organizational membership may strongly correlate.

The second type distinguished by Cole is *cognitive* particularism, which may be a necessary characteristic of science. Selecting someone for an academic position or a research grant because the applicant is in cognitive terms close to the decision makers makes sense, as those that decide most likely are convinced that their own discipline and own research line are very important. As at the research front uncertainty prevails, there are no 'objective' criteria to assess merit detached from preferences for specific approaches, topical fields and disciplines. But also here, reservations emerge, as systematic bias may come in, e.g., when those in the higher positions do not cover innovative and new interdisciplinary developments.

We distinguish various types of *social* particularism, based on different mechanisms. But the general characteristic is that all these mechanisms are based on some form of proximity between the involved persons. *Cronyism* literally points at an exchange relationship, e.g., one gives a research grant to someone and gets loyalty in return. *Nepotism* refers to giving favors based on family relations. But also other similarities between may play a role, such as gender: predominantly male decision makers select men more often than women for research grants. Such *gender bias* may be based on explicit opinions about differences between men and women, but it can also be unintentional and based on implicit gender stereotypes. Another form of particularism is *organizational proximity* which is bias emerging from organizational interest representation. A panel member may specifically support applicants that bring the grant (and the related prestige) to the panel member's university, even when the panel member does not have any personal connection to the applicant. Especially in case of prestigious and large grants, the organization's reputation benefits from receiving many of them. This paper focuses on the effects on the effect of particularism in the form of *organizational interest representation*.[3] That these variants of particularism should be avoided, is not contested. However, already Cole (1992) noticed that the boundaries are rather fuzzy. It could be the case that there is not only a friendship relation, but also a relation based on scientific reputation between an applicant and a panel member. Can we then empirically distinguish whether a decision was influenced by the one or by the other relation. Although Cole (1992) suggests that it is hardly possible to disentangle the two relations, we will try to do so in the empirical part of the paper.

*Cognitive proximity* is the other form of particularism. Panel members may support applicants from their own field or specialty, above better qualified applicants from other fields, even without knowing the applicants personally. As Cole already discussed, the problem comes up as whether this is particularism that should be avoided, or whether cognitive proximity is rooted in a basic characteristic of science: there is often disagreement between scientists about what the important problems are, and the about the best ways to answer those questions.

---

[3] This paper focuses on particularism in grant allocation, but particularism may also play a role in career decisions. The same mechanisms may play a role such as nepotism, gender bias or cognitive distance. It goes beyond the scope of this article to discuss the literature on bias in career decisions. Studies showing gender bias in career decisions are available for several countries, like – without claiming to be exhaustive - the Netherlands (Van den Brink 2010), Italy (Allesina 2011 – but contested by Abramo et al 2014a; Abramo et al 2014b, 2015) and e.g., Spain (Zinovyeva & Bagues 2012; 2015). But other studies argue the opposite.

Another relevant strand of theorizing comes from psychology. Thorngate et al. (2009) provide an interesting overview of what social psychology research has shown about merit and bias in small group decision making (Olbrecht & Bornmann 2010; Van Arensbergen at al. 2014). Overwhelming evidence shows that neutral decision making is very hard to achieve – if at all.

Research on particularism in grant allocation started back in the 1970s. Pfeffer et al. (1976), found that the distribution of NSF social science funding over universities correlated strongly with the number of panel members from those universities. This was stronger in fields with high uncertainty about what are the important research directions than in fields with low cognitive uncertainty. The same effect was found in other countries like Sweden (Sandstrom 2012), Korea (Jang et al. 2017), and Canada (Tamblyn et al 2018): "*One possible reason is that reviewers vote favorably for applicants from the same institution, even if they have never met them and would therefore not be in conflict*" (Tamblyn et al 2018).

A study of the NSF procedures found that reviewers and panel members did not favor proposals that came from applicants from their own state or region (Cole et al.1981). The well-known study by Wennerås and Wold (1997) suggested that grant decision making in a Swedish council suffered from considerable gender bias and nepotism. Ten years later, Sandström & Hällsten (2008) replicated this study and they found a similar level of nepotism, but no gender bias. Several studies indicate that panel members have higher success rates (Abrams 1991; Viner et al. 2004; Moed 2015), and Chinese studies showed the effect of connections to the research bureaucracy on getting research funding (Zhang et al., 2020). Van den Besselaar (2012) showed those that belong to the inner circle of a council – which is a much broader group than only panel members– have a significantly higher number of applications and an equally higher number of grants compared to applicants that are not involved in the (net)work of the council. This effect holds if one controls for past performance. This higher success (not success rates) may be based on being better informed about funding possibilities than those applicants that are more on a distance. When one is better informed, one can make better decisions when to apply (Van den Besselaar 2012; Bagues et al. 2019).

In order to avoid particularism, research councils have established conflict of interest (CoI) protocols, which generally means that if a panel member has a too close social or professional relation with an applicant, that panel member has to leave the panel meeting when the specific grant application is discussed, or even is not allowed in the panel whatsoever. Whether this is an efficient procedure has been questioned in a Canadian study (Gallo et al. 2016), and. despite the CoI-regulations, applicants with a link to a panel member seem to profit from that anyhow (Abdoul 2012).

An important problem, related to the distinction made by Cole (1992), is whether relations between applicants and panelists can and should be avoided at all, as one would expect the best (granted) applicants to have connections to panel members who should also be excellent researchers representing excellent research environments (Billig & Jacobsson 2009). This argument does not always holds, as panel members are not always excellent researchers (Abrams 1991; Sandstrom 2012), although in other cases they are – like in the case studied here (Van den Besselaar et al. 2015). This is a strong argument against interpreting a close relation between an applicant and panelists in terms of particularism, and we will come back to this argument in an empirical way, when analyzing our case below.

*The case.*

We investigate here the (2014) ERC starting grant. Applicants have to indicate in which research organization they will (when winning the grant) do the research. This means they can move with the grant from their affiliation when applying ('home organization' in ERC terminology) to another institution (the 'host organization'); of course they can also stay and then the host organization is the same as the home organization. Two mechanisms may be relevant. A panelist of the home affiliation of the applicant may be inclined to assess the application more negatively than justified by the quality, as he/she may see his/her organization losing a possible grant. A panelist from the host organization may, in contrast, be more positive than justified by the quality as his/her organization potentially wins a grant. These mechanisms cannot be understood as personal loyalty, but as representing organizational and through that one's own interest. As said, when a panel member is in the same organization (e.g. university) as the applicant there is a *conflict of interests* and the panelist or reviewer should leave the room when the application is discussed. Even if one assumes that this happens, it remains an open question whether it has the intended effect.

In this paper, we investigate the effect of *organizational proximity* on the success of grant applications, as an example of one of the various proximity relations that can exist between applicants on the one hand and peer reviewers and panel members on the other. This study differs from earlier studies as it is much broader in scope, and the case is one of the most prestigious grant schemes that currently exist, and with overall excellent scientists as panel members (Van den Besselaar & Sandström 2017). One would expect that if we find particularism in some form even here, it may exist everywhere.

## Data and method

We define organizational proximity in the following way: an applicant related to the same organization as at least one panel member. This relation can have two forms: An applicant either works at the same organization as one of the panelists, or a panelists is affiliated to the organization where the applicant – when receiving the grant – will move to. In line with the definition of the council, we define the organization at an aggregated level e.g. universities or national public research organizations like the Max Planck Gesellschaft in Germany, INRA in France, or CSIC in Spain. A first complicating factor is that research labs belonging to such PRO may be affiliated to a university – which seems increasingly to be the case. For example, several applicants employed at a CNRS institute also have a university affiliation and the specific location where the applicant works maybe a university institute. In other countries such double affiliations exist, e.g., in the Netherlands where the institutes of the Netherlands Royal Academy of Science and of the Netherlands research council NWO also have strong relations with universities. Here we use the affiliation mentioned by the applicants' CV and by the panelists' website. We use links at the level of the primary organizational affiliation, and not at the level of sub-units. This level of aggregation may be too high for studying nepotism, as this refers to a personal link between a panel member and an applicant. But, for studying organizational interest representation, this is adequate, as the personal relation does not need to play a role. A second complicating issue especially in France is the recent mergers within the university system, which makes attribution sometimes problematic. This was solved by searching on the web what the correct affiliation was in the period we study.

The following data were accessible for this case study. Panelists' names, countries and research field could be found on the council's website. Using this information, we identified the panelists' affiliation through searching the web for open CV data and home pages. A different strategy was followed for the applicant affiliation, as for these the council supplied the affiliation and the email address at the time of applying. In cases where the email address was a-specific (e.g., a Gmail address) we used the CV of the applicant. In some cases the CV was ambiguous, and that was solved by searching the web for the required information. In the case under study, applicants have to specify where they will use the grant: the so-called *host-institution*. Data about the intended host institute could be found in the applications as supplied by the council. Bibliometric data were retrieved from WoS and from Scopus. Data on earlier grants, and on the network of the applicants were extracted from the CVs of the applicants.

The host institution may be the same as their current affiliation (the *home institution*), but that is not necessarily the case. To compute proximity, we compared (i) the applicant's home institute with the panelists' home institutes and (ii) the applicants' host institute with the panelists' home institutes. By doing so, one can distinguish several forms of a *near-by panelist*:

- no near-by panelists (proximity-0)
- a panelist from the home institution of the applicant (proximity-1)
- a panelist from the applicant's intended host institution (proximity-2)
- a panelist from the home *and* from the host institution of the applicant (proximity-3)

The latter means in almost all cases that the applicants do not change institutions but plan to use the grant in the home institution. In only 3 cases, proximity 3 reflects a mobile applicant in combination with two different near-by panelists, one from the home and the other from the host institution. Proximity groups 1 and 2 are both very small. We report their size in the results section, but exclude them from the rest of the analysis.

Of the 3207 applicants of the 2014 ERC Starting Grant, 3030 signed the informed consent form. We checked whether the non-participating applicants (N=177) affect our findings, and that is not the case (Annex A1).

After comparing the success rate by proximity type, we analyze who profits from proximity. We do that at the country level, at the organization level, and at the individual level, where we specifically compare men and women applicants. In the analysis at the country level, we firstly exclude those countries that do not have any proximity relations in our sample. If there are no proximity relations, then the question of the effect of proximity relations on success rates makes no sense. Secondly, we also exclude countries with less than 50 applicants. We do this because in those cases, success rates change sharply with only one more or one less successful applicant.

Applicants from non-EU or non-associated countries (group-4 countries) are a special case, as prox-3 cannot occur: those applicants cannot stay in their home institution as it is outside of the EU and associated countries, and therefore these applicants have to move. This group of countries covers 143 applicants of which 23 successful – so the success rate (16.4 %) is higher than average. This set consists of two different subgroups, as the successful applicants come all from the U.S., Canada and Australia (and in 2014: Switzerland), whereas applicants from other non-EU countries are never successful. Furthermore, all women and most men among the successful applicants in this group are EU nationals returning to Europe, mostly from the U.S. Due to these peculiarities, and to the fact that prox-3, which is the dominant pattern in the rest of the data, does not exist for this group, we exclude group-4 countries from the current analysis.

Lastly, the Swiss situation warrants mentioning. In 2014, Swiss organizations could not act as a host for successful applicants as a consequence of the referendum that closed the Swiss borders for several groups of EU citizens. Although retroactively this was changed (in September 2014), this was too late for applicants to select a Swiss organization as host. As a consequence, Switzerland is treated here as a category 4 country. However, we did find two applicants who (according to the data we received from the ERC) were granted a StG, but moved to Switzerland. According to their homepages, they apparently got a different (replacement) grant through an ERC-SNF collaboration. We keep these cases in the analysis since we are interested in the decision making by the panels, and a panel did decide to award grants to these applicants.

### Research questions

We combine success rates with the proximity data, and then answer the following questions:

(i)   Is the success rate different for the different proximity-types distinguished above?

(ii)  Do the three domains (Life Sciences, Physics and Engineering, Social Sciences and Humanities) show the same pattern?

(iii) Who profits from proximity? Firstly, this will be analyzed at the individual level, in terms of gender differences. Secondly, it will be analyzed at the level of organizations: Does the ranking of universities correlates with profiting from organizational proximity? Thirdly, we investigate whether the host-countries differ in profiting from organizational proximity.

(iv) Can we explain different success rates also in other terms than particularism: Are those organizations that win most from proximity simply better, and therefore providing more panelists and at the same time attracting better and more successful applicants?

**Findings**

*Individual level*

Type-1 and type-2 proximity occur only a few times (Table 1), and therefore one more or one less case of these would change the effect as a percentage strongly. Therefore, we do not include prox-1 and prox-2 in the analysis. This is different for prox-3, where we have enough cases for further analysis. If an applicant makes clear in the application that he/she will remain in the same organization when receiving the grant, and there is a panel member from that organization, that panel member has an interest in the success of the applicant, and the data in Table 1 suggest that this interest may have an effect on the panel decisions, as the success rate of prox-3 cases is more than 40% higher than average. The question to be answered is why this is the case: are the prox-3 cases simply the better applicants, or is it the effect of interest representation? We address this issue below.

**Table 1**: Overall success rate versus success rate with a near-by panelist

| Proximity type | Applicants | Success | Success rate | vs all (11.69) |
|---|---|---|---|---|
| Proximity 0 | 2558 | 280 | 10.9% | 93.6% |
| Proximity 3 | 274 | 45 | 16.4% | 140.5% |
| All in sample | 2832 | 325 | 11.5% | 98.1% |
| Excluded groups | | | | |
| Proximity 1 | 31 | 3 | 9.7% | 82.8% |
| Proximity 2 | 22 | 2 | 9.1% | 77.7% |
| Group-4 countries | 145 | 23 | 15.9% | 135.7% |
| No consent | 177 | 22 | 12.4% | 106.3% |
| All applicants | 3207 | 375 | 11.7% | |

\* All cases except the non-response who did not get a grant, and except group-4 countries.

\*\* Numbers are too small for reliable interpretation.

The funding instrument under study has a two-step procedure, with a first selection where 75% of the applicants is rejected, and then a second selection where about half of the remaining applicants receive a grant. Table 2 shows that the nearby panelist effect is strongest in the first step, but also has some effect in Step 2.

**Table 2**: Success rate by near-by panelist (by Step)

| prox | total | to step 2 | SR | Prox-3 vs Prox-0 | granted | SR | Prox-3 vs Prox-0 |
|---|---|---|---|---|---|---|---|
| 0 | 2558 | 635 | 24.8% | | 280 | 44.1% | |
| 3 | 274 | 91 | 33.2% | 134% | 45 | 49.5% | 112% |
| Total | 2832 | 726 | 25.6% | | 325 | 44.8% | |

*Domain level*

The next question is whether domain differences occur, due to differences between disciplinary cultures. Table 3 summarizes the findings. In the life sciences, the pattern is the strongest and in line with the overall pattern: There, the success rate for the prox-3 group is *twice as high* compared to the success rate for the prox-0 group. Within Physics and Engineering Sciences, the effect of panel member proximity seems absent. Within Social Sciences and Humanities, panel member proximity shows a similar effect as in Life Sciences although the effect is smaller: a 40 % increase in success rate. As the number of prox-1 and prox-2 cases is low, and by domain even lower, they are not included in Table 3.

Table 3: Success rate with a near-by panelist versus overall success rate: domain differences

| Domain | ALL | | | Prox 0 | | | Prox 3 | | |
|---|---|---|---|---|---|---|---|---|---|
| | N | Success | success rate | N | Success | success rate | N | success | success rate |
| LS | 899 | 122 | 13.57% | 805 | 99 | 12.30% | 94 | 23 | 24.47% |
| PE | 1269 | 143 | 11.27% | 1144 | 128 | 11.19% | 125 | 15 | 12.00% |
| SH | 664 | 60 | 9.04% | 609 | 53 | 8.70% | 55 | 7 | 12.73% |
| ALL | 2832 | 325 | 11.48% | 2558 | 280 | 10.95% | 274 | 45 | 16.42% |

These results suggest that particularism does play a role in panel decisions in two of the three domains. However, it could also mean that the concentration of talent (panelists and applicants) is already substantial, and therefore, excellent applicants are in the same organizational environments as the excellent panelists. If concentration of talent would be the case, one would expect to find this more uniformly over all fields.[4] As we find significant differences between the fields, the findings seem to point at particularism, with substantial field differences.

It has been suggested that differences between the fields could be explained by competition: stiffer competition would lead to more unethical behavior such as cronyism in grant decisions. More specifically, the higher competition in life sciences could be the cause of the more substantial near-by panelist effect. However, our data do not support this explanation. In fact, competition in the LS domain is the lowest, demonstrated by the higher success rate in life sciences than in the two other domains. Another explanation relates to differences in the level of dependence and uncertainty between disciplines (Whitley 1980). Lower codification – meaning that there is less agreement of what is good science and in what direction a discipline is moving – would open up decision-making for nepotism. Testing this, however, needs analysis at an even lower level of aggregation and for this we currently lack the data.

## Who profits?

The next question is who profits from the different success rates for prox-3? We answer this question at the level of countries, of organizations, and of individuals.

### *Country differences*

To start with the first, one can distinguish between countries functioning as the *home*[5] country (where are the applicants working at the moment of applying) and countries functioning as *host* country (where applicants plan to spend their grant). As the hosts countries profit, we will focus on those.

For the descriptive statistics, we only include countries with more than 50 applicants. Using this set of countries[6], we find a strong correlation between the number of applicants and the number of successful applications as a host (r = 0.80), and a moderately strong correlation between the number of applicants and the number of proximity relations (r = 0.59). This is not surprising as more applicants indicate a bigger science system and thus more successful applicants and more panelists. There is also a strong correlation between the number of successful applicants and the number of proximity relations (r = 0.82). We however find no correlation (r = - 0.07) between the number of applicants and the success *rate* showing that the bigger systems are not outperforming the smaller ones.

---

[4] The issue of concentration of talent will be addressed below.

[5] This refers to residence, not to nationality.

[6] The following fourteen countries are included: Austria, Belgium, Denmark, Finland, France, Germany, Israel, Italy, Netherlands, Norway, Portugal, Spain, Sweden, UK.

In Table 4, we show the success percentages of the applicants by host countries, that may profit from the near-by panelist phenomenon. For this analysis, we only retain those countries that have at least 50 applicants and at least one proximity relation. Some countries show a much higher success rate for the group applicants with a near-by panel member[7] than their overall success rates, such as very strongly Finland, but also Sweden, Italy, UK, Germany, and Spain. This 'profit score' is in the last column of Table 4, and there one sees that e.g. the success rate of the UK within the group of applicants with a relation to a nearby panelist is about twice as high (1.8) as the overall success rate of applicants that have a UK based host organization. For Israel, France, and Denmark no the nearby panelist effect was found, and the Netherlands, Belgium, Austria, Portugal, and Norway show the opposite pattern and have a success rate lower than average for the near-by category.

**Table 4**: Near-by panelist advantage by country*

| Country | Number of applicants | Number success | Host success rate all (SR) | proximity (SRP) | SPR/SR |
|---|---|---|---|---|---|
| Finland | 103 | 7 | 6.8% | 28.6% | 4.2 |
| Sweden | 116 | 5 | 4.3% | 10.0% | 2.3 |
| Italy | 364 | 15 | 4.1% | 8.3% | 2.0 |
| UK | 539 | 62 | 11.5% | 20.6% | 1.8 |
| Germany | 395 | 65 | 16.5% | 27.3% | 1.7 |
| Spain | 269 | 22 | 8.2% | 11.1% | 1.4 |
| Israel | 85 | 22 | 25.9% | 28.6% | 1.1 |
| Denmark | 82 | 13 | 15.9% | 15.4% | 1.0 |
| France | 244 | 48 | 19.7% | 18.6% | 0.9 |
| Netherlands | 208 | 41 | 19.7% | 8.0% | 0.4 |
| Belgium | 101 | 10 | 9.9% | 0.0% | 0.0 |
| Austria | 82 | 13 | 15.9% | 0.0% | 0.0 |
| Portugal | 79 | 6 | 7.6% | 0.0% | 0.0 |
| Norway | 53 | 6 | 11.3% | 0.0% | 0.0 |

* Not included are countries without proximity relations or less than 50 applicants.

---

[7] This is not nationality or residency of the applicant but the country of the organization where someone applies the grant for.

Interestingly, there is a moderate negative correlation (r = - 0.36) between the profit a country has from proximity relations (SPR/SR) and the overall success rate of a country. This shows that to some extend the lower the country's overall success rate, the higher the benefit of proximity.

***Organizational differences***

Does the nearby panelist effect occur more in low ranked than in highly ranks organizations? Figure 1 shows that the median ranking of those granted with a nearby panelist relation is higher (median difference = 0.201, p = 0.156) than those granted without a nearby panelist relation, and also higher (0.256, p = 0.000) than the step-2 non-granted applicants. Also the mean values differ: 0.382 (p = 0.064) and 0.509 (p = 0.026) respectively. These findings suggest that high ranked organizations profit more of organizational proximity than lower ranked organizations do. This could support the argument that the nearby panelist effect is not so much particularism and bias, but the effect of preferential attachment: better organizations have more panelists and attract better applicants. We will test this below in the section on *organizational proximity and performance.*

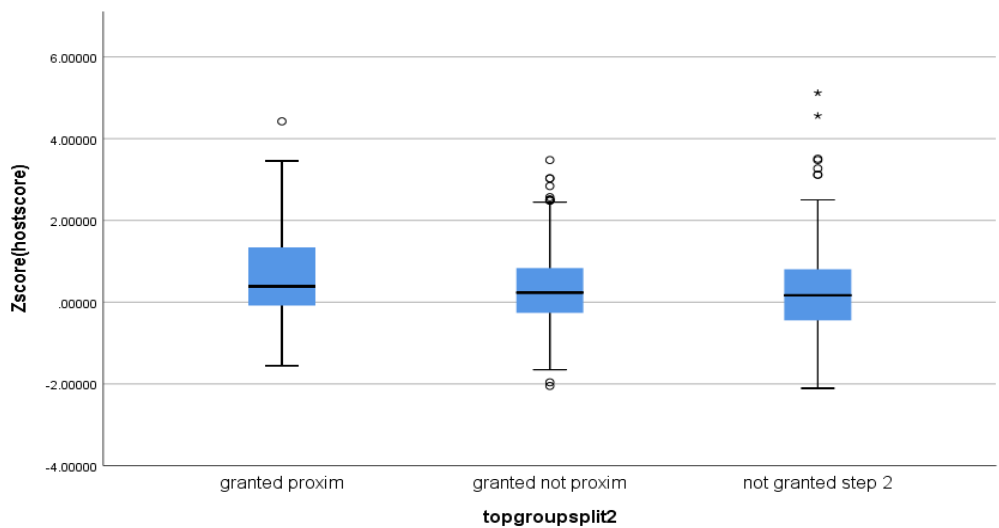

**Figure 1**: Median host organization ranking by proximity relation

### Gender differences

Table 5 shows that more men than women are involved in organizational proximity relations. Overall, male applicants are 1.4 times more likely than the female applicants to be in such proximity relation. This differs by domain, and for life sciences, for physics and engineering and for the social sciences and humanities the numbers are 1.37, 1.21 and 1.3 respectively.

Table 5: Distribution of proximity by applicants' gender and domain

|     |       | All  | Proximity | Share  |
|-----|-------|------|-----------|--------|
| LS  | men   | 481  | 63        | 13.10% |
|     | women | 324  | 31        | 9.57%  |
| PE  | men   | 858  | 98        | 11.42% |
|     | women | 286  | 27        | 9.44%  |
| SH  | men   | 318  | 36        | 11.32% |
|     | women | 291  | 19        | 6.53%  |
| All | men   | 1657 | 197       | 11.89% |
|     | women | 901  | 77        | 8.55%  |

Table 6 shows that when there is organizational proximity, men profit overall somewhat more from prox-3 then women do. Men have more often a prox-3 relation than women, and although for both men and women the chance of getting the grant increases when there is a proximity relation, the increase for men is higher: Organizational proximity seems to add to gender bias in grant allocation. However, the effect differs by domain. In the Life Sciences men profit slightly more than women from proximity, in the Social Sciences and Humanities men profit much more, and the pattern is exactly the opposite in Physics and Engineering, where women profit more than men do.

**Table 6**: Gender distribution of proximity by field (domain) and success.

| Field | Sex   | total success | | success with proximity | | Ratio* |
|-------|-------|------|--------|------|--------|--------|
| LS    | women | 41   | 11.55% | 6    | 19.35% | 1.68   |
|       | men   | 81   | 14.89% | 17   | 26.98% | 1.81   |
| PE    | women | 38   | 12.14% | 5    | 18.52% | 1.53   |
|       | men   | 105  | 10.98% | 10   | 10.20% | 0.93   |
| SH    | women | 28   | 9.03%  | 0    | 0.00%  | 0.00   |
|       | men   | 32   | 9.04%  | 7    | 19.44% | 2.15   |
| All   | women | 107  | 10.94% | 11   | 14.29% | 1.31   |
|       | men   | 218  | 11.76% | 34   | 17.26% | 1.47   |

*: ratio between the success rate with and without proximity

**Organizational proximity and performance**

Can the organizational proximity effect also be explained in a different way, without referring to particularism? The obvious alternative explanation is to take into account the performance of the host institutes and of the applicants. The hypothesis would be that the excellent applicants gravitate towards excellent organizations (Billig & Jacobsson 2009), and those excellent organizations are more likely to be present in the ERC panels than less excellent research organizations. In that case, the correlation between near-by panelist relations and grant success would be due to a confounding variable: excellence.

We test this by comparing the group of successful applicants with proximity-3 (*Granted-nearby*) with three other groups in terms of their scores on a few indicators, which are defined in Annex A2. We distinguish two types of indicators (Annex A3): (i) performance indicators, and (ii) prestige indicators. We compare the Granted-nearby with the group of granted applicants without a proximity relation (*Granted-other*), and with a group of excellent non-granted applicants, which is defined in two ways;

- the group that was not successful in the final phase of the procedure (*2Non-granted*)
- the non-granted with the highest performance score: the best of the rest (*BotR*). This BotR-group is selected per panel and is equally large as the set 2Non-granted. The 'best' is defined in terms of absolute impact.

The scores of these four groups are shown in Figure 2. The granted-nearby applicants score clearly better on the two *reputation* indicators: on *journal impact* of the journals in which they have published and on *network quality* in terms of the median ranking of the organizations found in the CVs of the applicants. The next highest on these two variables are the other granted applicants, and then the Step-2 non-granted. However, on the *performance* variable *total impact* the 'Best of the Rest' scores much higher than the other three groups, and on the *total grants* variable, the BotR score as high as the granted-nearby group. These findings suggest that one cannot explain the nearby panelist effect as preferential attachment based on excellent performance, as these granted-nearby applicants on average perform less than the three other groups. The granted-nearby group, however does score higher on the *reputation indicators*, so if it is a form of 'preferential attachment' and not of interest representation, it is reputation and not performance based.

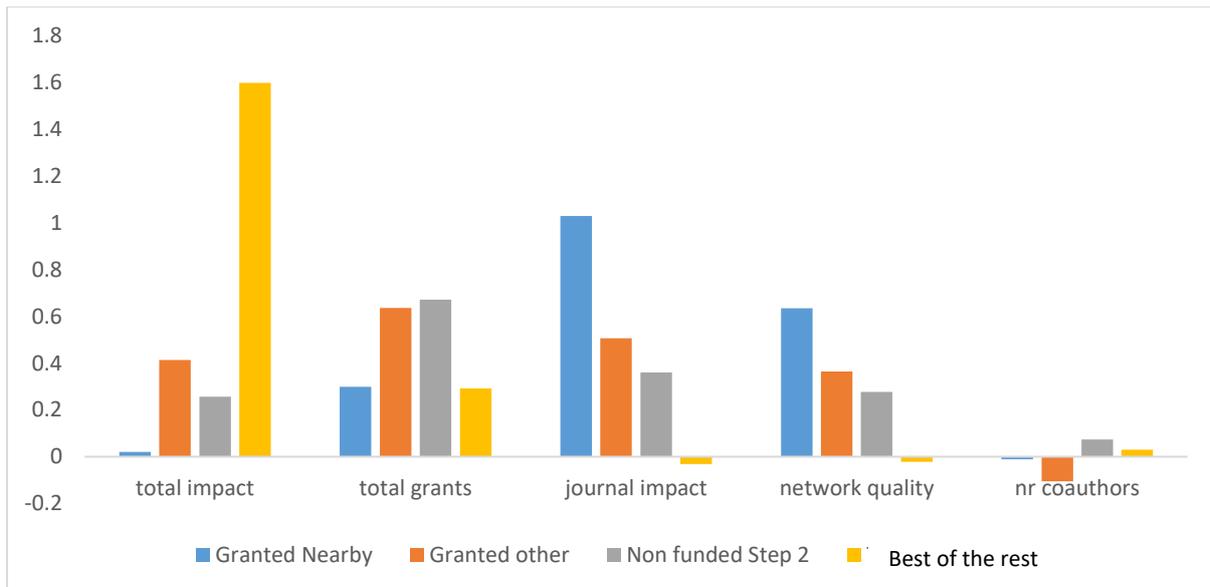

**Figure 2**: Scores on performance and reputation, four groups of applicants
All variables normalized at panel level

## Conclusions and further research

Our study suggests that having a nearby panel member does affect the grant decision process. Those with an organizational near-by panel member from the host institution have an overall much higher success rate than average, and the difference is substantial: 50% higher. We found that in PE the effect does not exist, but in LS the probability to get funded is twice as high with a nearby panelist relation. SH is in between these two.

Men profit somewhat more from a near-by panelist than women do. Interestingly, the exception is the Physics and Engineering domain, where women profit much more from a nearby panelist. This leads to interesting new questions: where do these domain differences come from? The field-based gender differences show a pattern that needs further exploration: The higher the share of women in a domain, the more men profit from the near-by panelist relation and vice versa. In the SSH with almost 50 % women applicants, men profit much more than women from proximity. In LS, with somewhat more than a third women applicants, men and women profit equally. And finally, within PE with about 25% women applicants, women profit much more than men do.

The hypothesis was formulated that the differences between the fields may be related to e.g. the level of competition. However, whereas the nearby panelist effect in the Life Sciences is much stronger than in the other fields, competition there is lower, as the success rate is higher in the LS than in the other domains. Also other field characteristics may play a role, such as the levels of uncertainty and dependence within a field (Whitley 2000). This could not be tested in the current study.

Also the country differences are substantial. Some countries profit much from the near-by panelist cases, in other countries there is no effect, and in again other countries the success rate of the near-by panelist cases is much lower than the average success rate. It remains an open question why these differences occurred. It may be a random pattern where countries profit in some years, but not in other years. If there would be a stable pattern, the question comes up as why some countries profit more than others from organizational proximity. Answering this would require repeating the study for more years.

Aside from the proximity effect at country level, we also showed that chances for non-EU nationals who do not already reside in the EU are zero. Understanding this observation would also need further research. Almost all successful applicants from outside the EU are EU citizens.

Finally, the question was asked as whether the observed advantage for applicants with a near by panelist is the result of interest representation, or of concentration of excellence. This alternative hypothesis that the near-by panelist advantage reflects the concentration of the most excellent researchers in the most excellent organizations was tested. We showed that the grantees with a near-by panelist have a much lower *performance level* than the grantees without a near-by panelist, and the difference is even larger with the highest performing non-grantees. On the other hand, the proximity-3 grantees scored substantially higher on *reputation* indicators. This suggests that reputation is (i) a vital asset in science, but (ii) not necessarily founded in performance. The conclusion of whether the nearby panelist effect is interest representation or a concentration of excellence depends on how one would understand excellence: as based on performance (as the authors of this paper are inclined to do) or as reputation.

Some limitations should be mentioned, which also point at directions for future research. (i) Although several variables are included to measure excellence (impact, earlier grants, top journals, quality of the network), excellence (and more generally quality) has more dimensions that may play a role in the grant decision making, such as independence (Van den Besselaar & Sandström 2019) which is explicitly mentioned by the council. (ii) Also other relational characteristics such as cognitive proximity (Sandström & Van den Besselaar 2018) may play a role. (iii) We could not observe the panels in their selection activities, but that would be crucial for understanding how the nearby panelist effect is produced. (iv) More work is needed on the methodological problem of identification of organizational affiliation and testing whether double affiliations have an effect on the results.

It is important for the science system, for the applicants, and also for the research councils to understand if and where particularism creeps into the grant selection procedures. Access to more data is for this a requirement. One would need data for more years and several funding instruments, to be able to investigate whether our findings can be generalized beyond the single case studied here.